\documentclass[aps,preprint,superscriptaddress,groupedaddress]{revtex4-1}

\usepackage[utf8]{inputenc}
\usepackage[french,english]{babel}
\usepackage[T1]{fontenc}
\usepackage[version=3]{mhchem} 
\usepackage{graphicx}
\usepackage{dcolumn}
\usepackage{bm}
\usepackage{xcolor}
\usepackage{subcaption}
\usepackage{amssymb}
\usepackage{amsmath}
\usepackage{epsfig}
\usepackage[colorlinks=false, pdfborder={0 0 0}]{hyperref}
\usepackage{cleveref}

\crefname{figure}{Figure}{Figures}
\crefname{equation}{Equation}{Equations}
\crefname{table}{Table}{Tables}

\graphicspath{{Figures/}}
\begin{document}
\title{Dual-scale turbulence in filamenting laser beams at high average power}

\author{Elise Schubert}
\affiliation{Universit\'e de Gen\`eve, GAP, Chemin de Pinchat 22, CH-1211 Geneva 4, Switzerland}

\author{Lorena de la Cruz}
\affiliation{Universit\'e de Gen\`eve, GAP, Chemin de Pinchat 22, CH-1211 Geneva 4, Switzerland}

\author{Denis Mongin}
\affiliation{Universit\'e de Gen\`eve, GAP, Chemin de Pinchat 22, CH-1211 Geneva 4, Switzerland}

\author{Sandro Klingebiel}
\affiliation{TRUMPF Scientific Lasers GmbH + Co. KG, Feringastraße 10A, 85774 Unterföhring, Munich, Germany}

\author{Marcel Schultze}
\affiliation{TRUMPF Scientific Lasers GmbH + Co. KG, Feringastraße 10A, 85774 Unterföhring, Munich, Germany}

\author{Thomas Metzger}
\affiliation{TRUMPF Scientific Lasers GmbH + Co. KG, Feringastraße 10A, 85774 Unterföhring, Munich, Germany}

\author{Knut Michel}
\affiliation{TRUMPF Scientific Lasers GmbH + Co. KG, Feringastraße 10A, 85774 Unterföhring, Munich, Germany}

\author{Jérôme Kasparian}
\affiliation{Universit\'e de Gen\`eve, GAP, Chemin de Pinchat 22, CH-1211 Geneva 4, Switzerland}
\email{jerome.kasparian@unige.ch}

\author{Jean-Pierre Wolf}
\affiliation{Universit\'e de Gen\`eve, GAP, Chemin de Pinchat 22, CH-1211 Geneva 4, Switzerland}

\date{\today}

\begin{abstract}
We investigate the self-induced turbulence of high repetition rate laser filaments over a wide range of average powers (1~mW to 100~W) and its sensitivity to external atmospheric turbulence. 
Although both externally-imposed and self-generated turbulences can have comparable magnitudes, they act on different temporal and spatial scales. While the former drives the shot-to-shot motion at the millisecond time scale, the latter acts on the 0.5~s scale. As a consequence, their effects are decoupled, preventing beam stabilization by the thermally-induced low-density channel produced by the laser filaments. 

\end{abstract}

\maketitle

\section{Introduction}

Propagating laser beams through turbulence is essential to many applications in the open atmosphere \cite{KaspaW2008}, including free-space communications~\cite{Kiasaleh2006,Belmonte1997,PolynPKRM2007}, remote sensing \cite{KaspaRMYSWBFAMSWW2003,GalveFIMIY2002,Gravel2004,Hemmer2011,Svanb}, the remote delivery of high-intensity for surface ablation \cite{Garcia2000, StelmRMYSKAWW2004, FujiiGMNN2006, TzortAG2006}, or weather modulation \cite{KaspaAAMMPRSSYMSWW2008a,HeninPRSHNVPSKWWW2011,Ju2014}.

Filamentation \cite{BraunKLDSM1995,ChinHLLTABKKS2005,CouaiM2007,BergeSNKW2007} is a self-guided propagation regime of high-intensity, ultrashort laser pulses. It relies on a dynamic balance between focusing by the Kerr effect and defocusing non-linearities~\cite{BejotKHLVHFLW2010a,Volkova2011,Richter2013}. These nonlinearities 
create transient refractive index gradients at least one order of magnitude larger than the random fluctuations induced by the strongest turbulence \cite{AckerMKYSW2006, SalamLSKW2007a}, so that filaments survive propagation through turbulence.

Recently, even much larger self-induced refractive index changes (up to 20\% \cite{Jhajj2014}) have been reported in the trail of laser filaments at repetition rates up to the kHz range, featuring sufficient average power to heat the air. The resulting depletion of the air density can partially persist until the next laser pulse, resulting in a cumulative effect \cite{Lahav2014, Cheng2013}. These thermal effects  induce a beam wandering~\cite{Yang2015}, enabling thermal self-action of beams previously restricted to long and energetic pulses, e.g., high-power CO$_2$ lasers~\cite{Kandidov2012}. 

Conversely, the channel of depleted air density left behind the filaments can also self-guide the beam~\cite{Lahav2014}, or be used in a filament array to stabilize and guide a  laser beam \cite{Rosenthal2014}. One could therefore expect that even more intense beams would self-stabilize, and even overcome the effect of externally imposed turbulence, with an increasing stability for increased incident average power and repetition rates.
 
We addressed this question by investigating the beam pointing stability of high-repetition rate laser filamenting beams over a wide range of average powers (1~mW to 100~W), with and without externally-imposed atmospheric turbulence. We show that the additional self-induced turbulence associated to high-repetition rate, high-power beam has a much longer time scale than the externally-imposed one. Consequently, they keep decoupled, preventing self-stabilizing of the beam. 

\section{Experimental setup}

\begin{figure}
	\centering
		\includegraphics[width=1.00\columnwidth]{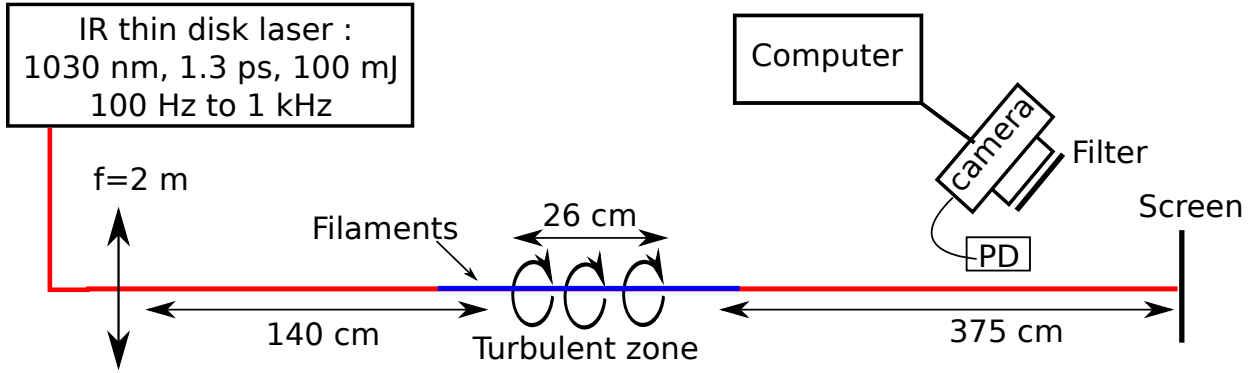}
			\caption{Experimental setup. Values correspond to the high-power beam}
		\label{fig:setup}
\end{figure} 

As sketched in Figure~\ref{fig:setup}, the experiment consisted in propagating the beam of an ultrashort laser beam through a turbulent hot air column generated by locally heating the air with a 26 cm long electric resistor. 

The turbulence of the hot air column was measured by propagating a low-power continuous-wave He:Ne laser on the same path as the main laser and measuring its beam pointing stability on a screen. The standard deviation $\sigma_\theta^2$ on the beam position was calculated, yielding the structure coefficient for the refractive index $C_n^2=\sigma_\theta^2 \left(2\omega\right)^{1/3}/2.91L$, where $L$ is the propagation distance within turbulence and $\omega$ is the width at $1/e$ of the intensity profile~\cite{AckerMKYSW2006, SalamLSKW2007a}. This value was estimated to be $1.5\times10^{-8}$~m$^{-2/3}$ with the resistor on, and $1.7\times10^{-11}$~m$^{-2/3}$ without it.

Two laser systems were used. The first one, hereafter denoted the moderate-power beam, delivered 82~fs pulses with an energy of 3~mJ at a wavelength of 800~nm. Its repetition rate was varied between 1 and 1000~Hz, corresponding to average powers of 1~mW to 1~W. The beam, of 13~mm initial diameter (at $1/e$) was slightly focused with an $f$~=~2~m lens, 3.5~cm above the electric resistor. It produced a 16~cm long filament at the waist, where the beam diameter was 0.4~mm. The turbulent region was placed at the most intense region of the filament, i.e., where the plasma acoustic wave was most intense. After a further 2.5~m, the beam was imaged on a Aluminium oxide screen and approximately 3000 single-shot images were recorded for each repetition rate, with a Phantom high speed camera (600 x 800 pixels) through an OD4 optical density and a long-pass 750~nm filter.

The second system~\cite{Klingebiel2015}, hereafter denoted the high-power beam, delivered 100~mJ, 1.3~ps pulses at~1030 nm, generating typically 3 -- 4 filaments of 50~cm length.
The repetition rate was varied between 100 and 1000~Hz, corresponding to average powers of 10 to 100~W: The use of longer pulses allows to reach much higher average powers for similar peak powers, hence similar levels of non-linearity. The geometrical configuration was identical to that of the moderate power beam, except for an initial beam diameter of 12~mm and the propagation distance between the turbulent region and the aluminium oxide screen, namely  3.75~m. In each experimental condition, approximately 1900 single-shot images were recorded at 50~ms time interval using a PixelInk PL-B761U CCD camera with 480~x~752~pixels, through a Schott BG7 filter. 
 
The beam position was analyzed as follows. Each image was thresholded at 10\% of the maximum intensity to isolate the beam from the background. The center of mass of this area provided us with the beam position.
The position of the filament in the moderate-power beam was determined using a higher threshold (87\% of the maximum intensity). The beam wandering was characterized by two parameters: its magnitude, defined as the two-dimensional standard deviation $\sigma_\theta$ of the pointing direction, and its instantaneous speed $v_\textrm{w}$, defined by the displacement of the beam position between two successive images, divided by the corresponding time interval.

As the phase mask induced in the air by turbulence evolve slowly as compared to the beam repetition rate, the beam wandering speed can be acurately resolved. As a consequence, if the time interval between the laser pulses is sufficiently short, their successive positions on the screen will display some correlation. We investigated the corresponding time constant by calculating the autocorrelation function of these positions and defined the corresponding correlation time as the decay time (at $1/e$) of this beam position autocorrelation function. 

We finally quantified the deviation of the beam profile from its initial circular shape by calculating the two-dimensional second moment of the thresholded image.

\section{Results and discussion}

\begin{figure}[tb]
	\centering
	\includegraphics[width=1\columnwidth]{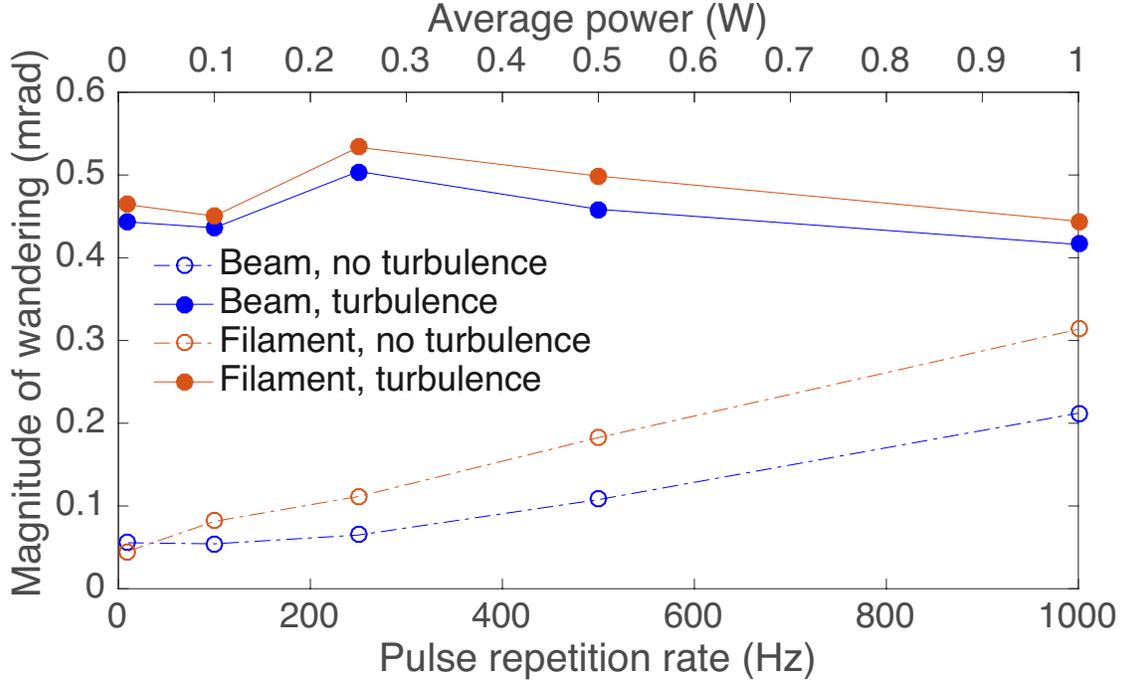}
	\caption{Wandering of the single filament and of the whole moderate-power beam propagating through a turbulent region}
	\label{fig:pointe_filament}
\end{figure}

The 1~mJ pulse energy of the moderate-power laser ensures that the beam constantly produced one single filament, which was sufficiently strong to survive the propagation through the turbulent region. Without external turbulence, the wandering of the filament  increases with the repetition rate and thus the average power (Figure~\ref{fig:pointe_filament}), due to the turbulence induced by the energy deposited by the filaments in the air~\cite{Lahav2014, Cheng2013,Yang2015}. Accordingly, the standard deviation of the filament pointing rises from 50~$\mu$rad at 10~Hz to 310~$\mu$rad at 1~kHz.
The self-induced turbulence extends beyond the filament itself: it increases the wandering of the whole beam, although to a lesser extent ($\sigma_\theta = 210\ \mu$rad at 1~kHz).
 
When imposing an external turbulence, both the filament and the whole beam are randomly deflected,  further increasing their wandering. The standard deviation rises to $\sim450\ \mu$rad, independent of the repetition rate. This independence shows that the effects of the self-induced turbulence do not influence those of the externally-imposed one, that dominates the beam propagation when present. As a consequence, no self-stabilization, nor self-destabilization is observed.

\begin{figure}[tb]
	\centering
	\includegraphics[width=1\columnwidth]{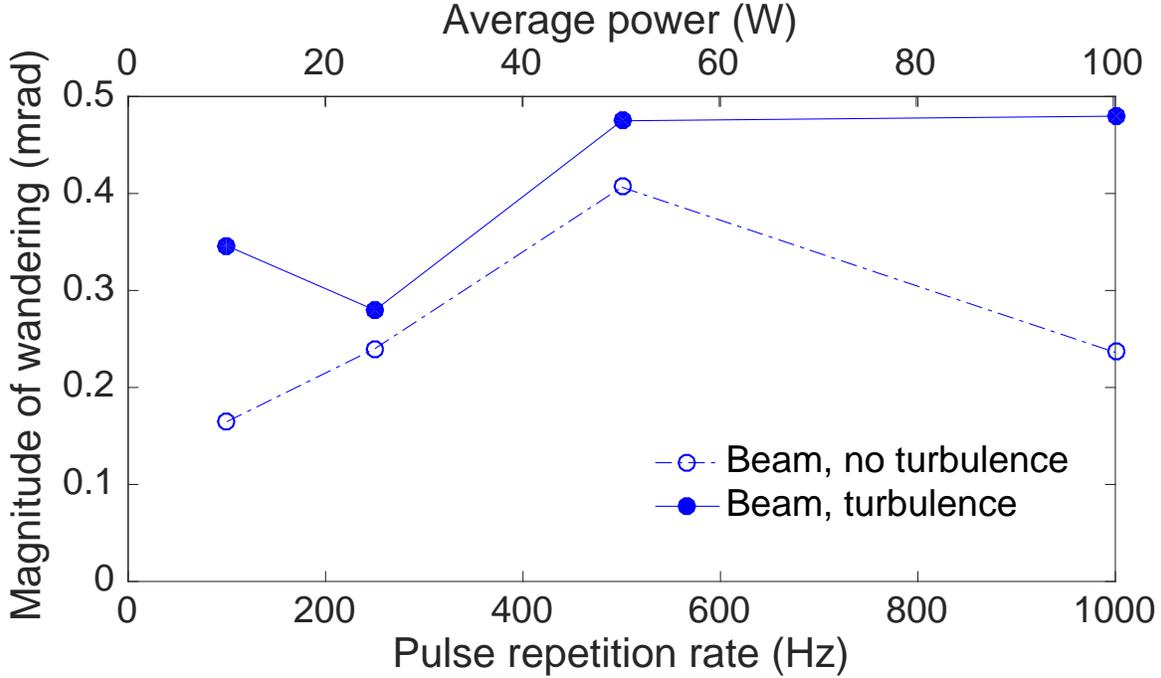}
	\caption{Wandering of the high-power laser beam beam propagating through a turbulent region}
	\label{fig:stability}
\end{figure}

Surprisingly, in a 100-fold higher power range, the self-induced turbulence is only slightly stronger. This can be understood by considering that the energy deposition is much less efficient with our high-power laser system. Indeed, Houard \emph{et al.} measured the energy losses for laser pulses identical (resp. similar) to those produced by the high-power (resp. moderate-power) system. These losses are 10 times lower (2\% \emph{vs.} 20\%) for 1030~nm, 1.5~ps pulses than for 800~nm, 100~fs ones~\cite{Houard2016}. Furthermore, the formation of a pattern of several filaments may also result in a more homogeneous energy deposition, limiting the associated refractive index gradients.
The increase of the beam wandering in externally-imposed turbulence is also influenced  by the strong intrinsic dependence of the laser pointing stability. 
Without external turbulence, the beam pointing stability of the high-power laser system itself is optimized for operation at 1000~Hz repetition rate. At intermediate repetition rates (500~Hz), the thermal lensing or other processes inside the amplifiers degrade the pointing stability.

With the externally-imposed turbulence, the beam wandering is larger, with an angular standard deviation between 350 and 480~$\mu$rad (Figure \ref{fig:stability}).
Furthermore, this wandering does not seem to depend on the repetition rate: no beam self-stabilization occurs when the repetition rate increases, even at the 100~W average power level.

\begin{figure}[tb]
	\centering
	\includegraphics[width=1.00\columnwidth]{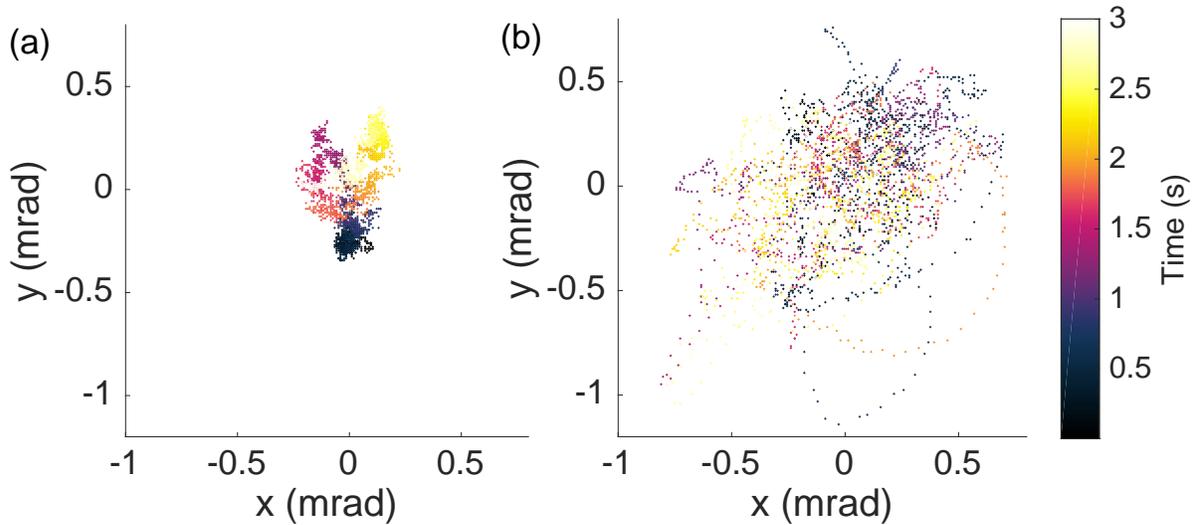}
	\caption{Trajectory of the moderate-power beam pointing at a repetition rate of 1~kHz, (a) without and (b) with externally-imposed turbulence. See also Supplementary Movies 1 and 2, respectively corresponding to Panels a and b}
	\label{fig:positions}
\end{figure}

The fact that the self-induced thermal effects do not affect the wandering through the externally-induced turbulence can be understood by observing their respective influences on the beam trajectory on the screen, displayed on \Cref{fig:positions}.
Without externally-imposed turbulence, i.e. under the control of the self-induced turbulence, the beam stays confined for typically 0.5~s within a cluster with dimensions of approximately 100~$\mu$rad. Between these confinement times, it experiences rare long-distance (sub-mrad) jumps to a new cluster position (\Cref{fig:positions}a, and Supplementary Movie 1). 
Accordingly, the beam position displays correlation times up to 110~pulses (0.1~s) at 1~kHz.

\begin{figure}[tb]
	\centering
		\includegraphics[width=1.00\columnwidth]{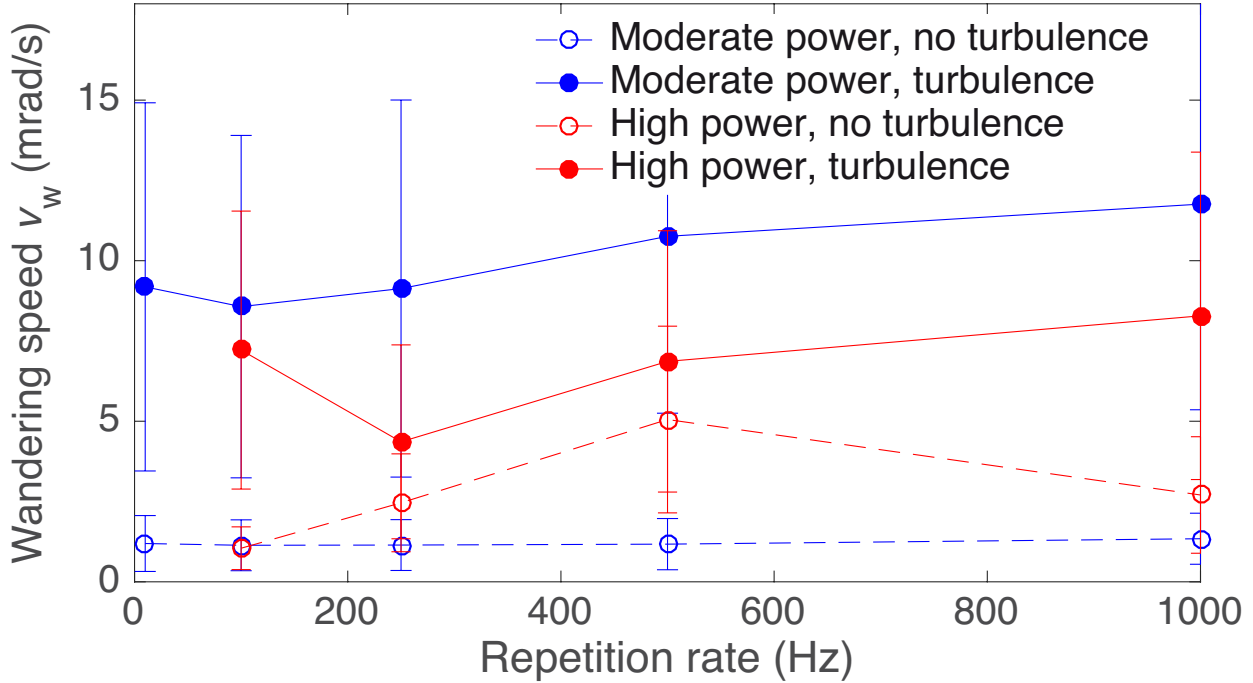}
			\caption{Instantaneous wandering speed $v_\textrm{w}$ of the laser beam}
	\label{fig:step}
\end{figure}

As the jumps between clusters are larger than the typical cluster size, they govern the overall magnitude of the beam wandering. The increase of the beam wandering for higher repetition rates is therefore the signature of longer jumps. However, these jumps are sufficiently rare to have a negligible influence on the average wandering speed (See dotted lines in \Cref{fig:step}), which is rather representative of the beam motion within the clusters.

Externally-imposed turbulence substantially increases the shot-to-shot wandering of the beam within the clusters, so that the latter grow in size and widely overlap each other (\Cref{fig:positions}b and Supplementary Movie 2). Consequently, both the speed (See solid lines in \Cref{fig:step}) and the magnitude of the wandering (Figures~\ref{fig:pointe_filament} and \ref{fig:stability}) are governed by the size of the clusters and keep independent of the jumps between them, hence of the self-induced turbulence. 

In summary, the external turbulence governs the shot-to-shot motion of the beam within the clusters, while the self-induced one is responsible for the jumps from cluster to cluster at a much longer time scale of 0.5~s. 
This behavior is observed not only for the whole beam, but also for the individual filaments. In the latter case, the wandering speed is slightly higher, by $\sim$1~mrad/s.

We attribute the longer time scale of the self-induced turbulence to smoother and wider refractive index gradients produced by the local convection, as compared with the externally-imposed one, the vorticity of which has more space to develop during the transport of the corresponding air mass to the laser beam location.

This dual time scale may explain the apparent discrepancy of our results with those of Yang \emph{et al.}~\cite{Yang2015}, who measure a structure parameter for the refractive index $C_n^2$ typically two orders of magnitude smaller than in our work ($C_n^2 \approx 5\times10^{-9}$~m$^{-2/3}$) for  2~mJ, 45~fs pulses, comparable to our moderate-power system. 
However, they measured the fluctuation of the beam position over only 20 pulses (i.e., 20~ms). Therefore, they focus on the the short-scale wandering within a cluster, i.e., to the effect of the residual external turbulence in the laboratory. 
Conversely, several seconds (3000 pulses in our setup correspond to 3~s) are necessary to get access to the full excursion of the beam pointing controlled by the self-induced turbulence.

Finally, we observe that neither the self-induced nor the externally imposed turbulences significantly affect the beam shape and its filamentation. 
The beam profile, characterized as described above by its two-dimensional second moment, is almost unchanged by both turbulence sources.
This is presumably due to the fact that the beam diameter does not exceed the inner scale of the turbulence ($l_0 \sim 1$~mm) ~\cite{Meyzonnette2001}, so that the refractive index  gradient is homogeneous over the beam profile.

The filament formation process is also unaffected by the turbulence. At moderate power, the filament and the full beam display very similar trajectories (Figure~\ref{fig:pointe_filament}). Indeed, the correlation coefficient between their positions ranges between 0.96 and 0.98 at all investigated repetition rates. 

At full power (100~mJ, 100~W average power, 76~GW peak power, i.e., $\sim$15~$P_\mathrm{cr}$),  80 – 100 \% of the filaments  survive the propagation through the turbulent region. This survival probability does not depend significantly on the repetition rate. Such robustness of filaments is consistent with prior results at lower repetition rate and average power (22.5~Hz, 14~mW)~\cite{AckerMKYSW2006}.

\section{Conclusion}

As a conclusion, we investigated the respective effects of externally-imposed and self-generated turbulence on high-repetition rate, ultrashort-pulse lasers with average powers ranging from 1~mW up to 100~W. 
Externally-imposed and  self-induced turbulences display very different time constants, in the millisecond and sub-second ranges, respectively. 

Furthermore, the self-induced thermal effects do not affect the effect of the the externally-imposed turbulence, so that they induce no beam self-stabilization in the investigated range of powers. However, the long time scale of the self-induced turbulence at high repetition rates may make it easier to compensate using adaptive closed-loop techniques, potentially helping  propagating high-power, high repetition rate pulses through perturbed atmospheres.

\textbf{Acknowledgements}. We acknowledge financial support from the ERC advanced grant « Filatmo ». Technical support by M. Moret was highly appreciated. 

\bibliography{Biblio2}

\end{document}